\documentclass[manuscript,screen]{acmart}

\usepackage{mathtools}
\usepackage{algorithm}
\usepackage{algpseudocode}
\usepackage{xcolor}
\usepackage{subcaption}
\usepackage[normalem]{ulem}
 \newcommand{\revtext}[1]{\textcolor{black}{#1}}

\usepackage{enumitem}
\hypersetup{colorlinks=true, urlcolor=blue, linkcolor=black, citecolor=black}

\newcommand{\E}{\mathbb{E}}

\newcommand{\I}{\mathbb{I}}

\newcommand{\G}{\mathcal{G}} 
\newcommand{\X}{\mathcal{X}} 
\newcommand{\Z}{\mathcal{Z}} 
\newcommand{\Y}{\mathcal{Y}} 

\AtBeginDocument{%
  \providecommand\BibTeX{{%
    \normalfont B\kern-0.5em{\scshape i\kern-0.25em b}\kern-0.8em\TeX}}}

\newif\ifdraft
\drafttrue

\begin{document}

\setcopyright{none}

\title[Multi-party Computation Protocols for Post-Market Fairness Monitoring]{Co-designing for Compliance: Multi-party Computation Protocols for Post-Market Fairness Monitoring in Algorithmic Hiring}

\author{Changyang He}
\authornote{This work was primarily conducted during the author's stay at the Max Planck Institute for Security and Privacy.}
\affiliation{%
  \institution{Harbin Institute of Technology, Shenzhen}
  \city{Shenzhen}
  \country{China}
}
\affiliation{%
  \institution{Max Planck Institute for Security and Privacy}
  \city{Bochum}
  \country{Germany}
}
\email{hechangy@hit.edu.cn}

\author{Nina Baranowska}
\affiliation{%
  \institution{Leiden University}
  \city{Leiden}
  \country{Netherlands‌}
}
\email{n.n.baranowska@law.leidenuniv.nl}

\author{Josu Andoni Eguíluz Castañeira}
\affiliation{%
  \institution{Universitat Pompeu Fabra}
  \city{Barcelona}
  \country{Spain}
}
\affiliation{%
  \institution{InfoJobs}
  \city{Barcelona}
  \country{Spain}
}
\email{josu.eguiluz@adevinta.com}

\author{Guillem Escriba}
\affiliation{%
  \institution{InfoJobs}
  \city{Barcelona}
  \country{Spain}
}
\email{guillem.escriba@adevinta.com}

\author{Matthias Jüntgen}
\affiliation{%
  \institution{Ruhr University Bochum}
  \city{Bochum}
  \country{Germany}
}
\email{matthias.juentgen@rub.de}

\author{Anna Via}
\affiliation{%
  \institution{InfoJobs}
  \city{Barcelona}
  \country{Spain}
}
\email{anna.via@adevinta.com}

\author{Frederik Zuiderveen Borgesius}
\affiliation{%
  \institution{Radboud University}
  \city{Nijmegen}
  \country{Netherlands}
}
\email{frederikzb@cs.ru.nl}

\author{Asia J. Biega}
\authornote{Corresponding author.}
\affiliation{%
  \institution{Max Planck Institute for Security and Privacy}
  \city{Bochum}
  \country{Germany}
}
\email{asia.biega@mpi-sp.org}

\renewcommand{\shortauthors}{He et al.}

\begin{abstract}
Post-market fairness monitoring is now mandated to ensure fairness and accountability for high-risk employment AI systems under emerging regulations such as the EU AI Act. However, effective fairness monitoring often requires access to sensitive personal data, which is subject to strict legal protections under data protection law. Multi-party computation (MPC) offers a promising technical foundation for compliant post-market fairness monitoring, enabling the secure computation of fairness metrics without revealing sensitive attributes. Despite growing technical interest, the operationalization of MPC-based fairness monitoring in real-world hiring contexts under concrete legal, industrial, and usability constraints remains unknown. This work addresses this gap through a co-design approach integrating technical, legal, and industrial expertise. We identify practical design requirements for MPC-based fairness monitoring, develop an end-to-end, legally compliant protocol spanning the full data lifecycle, and empirically validate it in a large-scale industrial setting. Our findings provide actionable design insights as well as legal and industrial implications for deploying MPC-based post-market fairness monitoring in algorithmic hiring systems.

\end{abstract}



\maketitle

\section{Introduction}


Algorithmic hiring is increasingly central to human resource management due to its efficiency and scalability~\cite{lashkari2023finding,fabris2025fairness}. However, extensive evidence indicates that such systems can perpetuate or even amplify discrimination, bias, and inequality~\cite{sanchez2020does,raghavan2020mitigating,fabris2025fairness}. In addition to training and validating fair hiring algorithms before deployment, \textbf{post-market fairness monitoring} has become essential for ensuring accountability and mitigating unfair outcomes in real-world hiring contexts~\cite{mokander2022conformity}, and is now mandated by technology regulations~\cite{EU_AI_Act_Article_72}. In particular, the European Union's AI Act classifies employment-related systems as high-risk~\cite{edwards2021eu}, and obliges providers to establish and document a post-market monitoring system~\cite{EU_AI_Act_Article_72}. This post-market monitoring system shall actively and systematically collect, document, and analyze relevant data to evaluate the continuous compliance of AI systems throughout their lifetime~\cite{EU_AI_Act_Article_72}, including the detection and mitigation of possible algorithmic biases~\cite{EU_AI_Act_Article_10}.

Nonetheless, the mandate of fairness monitoring introduces a fundamental tension: reliable fairness measurement often depends on access to sensitive personal data such as ethnicity and sexual orientation~\cite{fabris2025fairness,raghavan2020mitigating,mehrabi2021survey}, yet these sensitive attributes are subject to strict legal protections. European Union's General Data Protection Regulation (GDPR) designates such information as special categories of data and prohibits their processing by default~\cite{voigt2017eu,hoofnagle2019european}. On this note, the EU AI Act explicitly allows the use of special categories of personal data for bias detection and correction only in the context of training, validation, and testing of high-risk AI systems, under strict necessity and safeguards~\cite{EU_AI_Act_Article_10}; However, the AI Act does not provide an explicit legal basis for the use of such data for post-deployment fairness monitoring~\cite{EU_AI_Act_Article_72}.

Privacy-preserving techniques \cite{glerean2025secureai} such as differential privacy (DP) and multi-party computation (MPC) have been proposed as viable technical foundations for compliant fairness monitoring, allowing post-market assessments without direct disclosure of sensitive attributes~\cite{goldreich1998secure,helminger2022multi}. Among these approaches, MPC is particularly promising because it preserves data fidelity by avoiding noise injection~\cite{zaccour2025access}, which can be advantageous for fairness monitoring in hiring contexts where sample sizes are often limited. In such settings, MPC-based fairness monitoring typically distributes sensitive attributes across multiple parties, who jointly compute group-level fairness metrics through secure protocols without revealing individual-level information to any party~\cite{pentyala2022privfair,pentyala2022privfairfl,kilbertus2018blind}. However, existing work largely remains at the level of conceptual or technical feasibility, with limited consideration of how MPC-based fairness monitoring can be operationalized in practice under concrete legal compliance requirements, real-world industrial constraints, and end-to-end data lifecycle considerations.

To address this gap, this work aims to develop MPC protocols for fairness monitoring that cater to real-world needs. We adopt a co-design approach that integrates computational, legal, and industrial expertise in algorithmic hiring. Through iterative discussions among the research team, we first capture real-world legal and industrial design requirements for MPC fairness monitoring protocols, covering aspects such as \textit{data and privacy requirements}, \textit{fairness requirements}, \textit{usability requirements}, and \textit{feasibility requirements}. We then design the framework for an MPC-based fairness monitoring protocol, spanning the different stages of a system's lifecycle, from data collection and encryption to fairness computation and front-end presentation. We implemented and empirically validated the protocol using a fairness monitoring dashboard within a large-scale, real-world industrial environment. Based on the study, we derive legal and industrial implications of MPC-based post-market fairness monitoring protocols in algorithmic hiring.

In summary, this work makes the following contributions: (1) We \emph{identify the practical requirements} for MPC-based fairness monitoring in algorithmic hiring, grounded in real-world legal compliance obligations and industrial deployment constraints. (2) We \emph{design} a legally compliant and deployment-oriented MPC fairness monitoring protocol that supports end-to-end fairness monitoring across the data lifecycle, explicitly accounting for industry practices and regulatory requirements.\footnote{\revtext{Our open-source implementation is available in the monitoring package of the findhrAPI library: https://github.com/findhr/findhrAPI}} (3) Through \emph{empirical validation in a real-world industrial setting}, we derive concrete design insights and practical lessons for implementing MPC-based post-market fairness monitoring in algorithmic hiring systems.

\section{Related Work}

\noindent \textbf{Post-deployment fairness monitoring necessitates privacy-preserving protocols.} Although mandated by the EU AI Act, post-market fairness monitoring in hiring systems faces an inherent tension between the need for reliable fairness assessments, which typically require access to sensitive attributes, and the stringent privacy constraints governing the collection and use of such data~\cite{holstein2019improving,zaccour2025access,veale2017fairer,chang2021privacy,stadler2022search}. In response, the algorithmic fairness and privacy communities have developed a range of privacy-preserving protocols for post-deployment fairness monitoring~\cite{zaccour2025access,chang2021privacy,islam2023differential}. Typical examples are Differential Privacy (DP) and Multi-Party Computation (MPC).

\noindent \textbf{Differential Privacy (DP).} Differential Privacy is a formal privacy framework that bounds the influence of any single individual on released statistics, thereby limiting disclosure risk~\cite{dwork2006differential}. In fairness monitoring, DP is typically implemented by adding calibrated noise to aggregate statistics such as confusion matrices or group-level fairness metrics to enable privacy-preserving fairness assessment~\cite{zaccour2025access,chang2021privacy,jagielski2019differentially}. Nonetheless, the injected noise reduces statistical fidelity, and smaller datasets or stricter privacy budgets might lead to unreliable estimates~\cite{zaccour2025access}.

\noindent \textbf{Multi-Party Computation (MPC).} Consider a system with $n$ parties $P_1,\ldots,P_n$, where each party $P_i$ holds a private input $x_i$. Given a function $f$, a secure multi-party computation (MPC) protocol enables the parties to jointly compute $y = f(x_1,\ldots,x_n)$ such that no party $P_j$ learns any information about another party's input $x_i$ for $i \neq j$, beyond what is implied by the output $y$~\cite{goldreich1998secure}. MPC is widely adopted in privacy-sensitive collaborative settings, such as secure data analytics~\cite{sahinbas2023secure,zhao2025libertas}, federated machine learning~\cite{mugunthan2019smpai}, and cross-organizational statistical analysis~\cite{elkoumy2020secure}, where mutually distrustful parties could compute joint functions over confidential data. For instance, MPC enables analytics on sensitive medical data that complies with European privacy legislation~\cite{veeningen2018enabling}. Therefore, MPC provides a promising approach to privacy-preserving fairness monitoring by allowing parties to jointly compute fairness metrics without disclosing users' sensitive attributes~\cite{helminger2022multi,espiritu2024synq,qin2019usability}. As a notable example, the Boston Women's Workforce Council employs MPC to evaluate wage disparities across gender and racial groups while preserving the confidentiality of each party's private data~\cite{lapets2018accessible,BWWC,qin2019usability}. In contrast to DP and synthetic data approaches, which rely on noise injection or distributional approximations, MPC preserves higher fidelity of results by enabling fairness metrics to be computed directly over encrypted data, while still providing strong cryptographic privacy guarantees.

\noindent \textbf{Challenges of MPC protocol implementation and our contributions.} Although MPC enables higher-fidelity computation of fairness metrics while providing strong privacy guarantees, its practical deployment is accompanied by significant challenges. For instance, a common MPC instantiation is secure two-party computation, where the deployer of the AI-based hiring system collaborates with a trusted third party (TTP) to jointly compute fairness metrics over encrypted data; however, it remains a theoretical hypothesis. The implementation details, from data collection, data storage, fairness computation, to result presentation, are still unclear, and how to align these implementation details with legal and industrial requirements remains an unexplored yet critical problem. To fill this gap, we adopt a co-design approach to combine technical, legal, and industrial expertise, informing practical MPC fairness monitoring protocols in algorithmic hiring.

\section{Co-Design Approach}

To integrate legal compliance requirements, state-of-the-art academic research on privacy-preserving and fair algorithm design, and practical deployment considerations from the industry, this work involved a three-party interdisciplinary collaboration based in Europe. The collaboration brought together (1) a computer science research team focusing on privacy-enhancing technologies and responsible AI, (2) a legal research team specializing in privacy, data protection, and discrimination law, and (3) an industry team that deploys a large-scale algorithmic recruitment system, one of the leading employment platforms in Europe that has over 10 million candidates and 60,000 active companies.

First, the legal and industry parties independently developed the design requirements. The legal research team, drawing on laws including the AI Act, the GDPR, and non-discrimination laws, organized the legal definitions and requirements for post-market fairness monitoring. They further examined potential tensions between fairness monitoring and data protection and developed a set of legal requirements for fairness monitoring, as detailed in Section \ref{sec:legal-requirements}. The industry team reflected on the real-world objectives and challenges of fairness monitoring through internal discussions among different stakeholders (including developers, product managers, and the internal legal team). Based on these insights, they further refined and developed the industry requirements for fairness monitoring, as shown in Section \ref{IndustryRequirements}.

Next, to turn the requirements into a protocol design, the research team adopted an iterative multi-stakeholder co-design approach. \revtext{Between 2024 and 2025, the computer science and industry teams collaborated to discuss design requirements and develop feasible fairness monitoring solutions, supported by the legal team’s compliance-oriented input.} This co-design process resulted in the MPC post-market fairness monitoring protocol presented in Section \ref{systemModel}. Our open-source implementation of the protocol\footnote{\revtext{The MPC protocol development is based on the MPyC library: https://mpyc.readthedocs.io/en/latest/mpyc.html}} is available in the \textit{(monitoring)} package of the findhrAPI library\footnote{\revtext{https://github.com/findhr/findhrAPI}}.


Next, the industry team implemented the protocol and a fairness monitoring dashboard tailored to their employment platform, following the development of usability considerations detailed in Section \ref{IndustrialDashboard}. The validation in an industrial environment has also led to minor revisions of the protocol. Finally, all parties reflected on lessons learned from this co-design work and summarized the legal and industrial implications in Section \ref{lessons}.

\section{Design Requirements}

\subsection{Legal Requirements for Fairness Monitoring}
\label{sec:legal-requirements}

\noindent \textbf{The AI Act.}The AI Act (AIA)~\cite{act2024eu}, which is the EU regulation that sets common rules for the development and use of AI systems, requires providers (generally organizations that develop AI systems) of high-risk systems to \textbf {detect and mitigate discriminatory bias} during development and to continuously monitor the system when it operates on the market. In addition, Article 72 AIA~\cite{EU_AI_Act_Article_72} imposes a general obligation on providers to conduct \textbf{post-market monitoring} to ensure that AI systems continue to meet legal and technical requirements and work as intended. Checking the system for potential discriminatory outputs is one of the monitoring elements. However, for now, the monitoring obligation does not specify the monitoring design. The AI Act merely states that monitoring must actively and systematically collect, document, and analyse relevant data that can come from deployers (e.g., recruiters) and other sources to assess AI system performance throughout its lifecycle. The results of post-market monitoring should help identify, analyze, evaluate, and mitigate discrimination risk and contribute to the broader risk management system, required by Art. 9 AIA~\cite{EU_AI_Act_Article_9} and understood as a continuous lifecycle process with regular review and updates. 


In practice, hiring companies can act as both providers and deployers~\cite{fabiano2025subject}. Considering this potential dual role, a fairness monitoring protocol should be designed with industrial feasibility and practical usability in mind, and should incorporate the deployer's perspective, which takes into account the realities faced by recruiters and hiring companies.


\noindent \textbf{Data protection law.} The way how companies can monitor the performance of AI systems depends to a certain extent on the General Data Protection Regulation (GDPR)~\cite{voigt2017eu}. This is because assessing if an AI system disproportionately disadvantages protected groups relies on fairness metrics, which, in turn, require information on individuals’ attributes (\textbf{`personal data'} in legal terms). First, the GDPR determines whether a company can legally process (collect, access, use, store, etc.) personal data at all. Second, once processing is allowed, the GDPR imposes requirements, including data minimization, purpose limitation, storage limitation, and privacy by design and by default, which are also important for fairness monitoring. The proposed protocol's design incorporates these legal requirements. 

The GDPR classifies some of the attributes necessary for fairness monitoring, including ethnicity, religion, sexual orientation, and health, as \textbf{‘special categories of personal data’} under Art. 9 GDPR, also called ‘sensitive data’~\cite{EU_GDPR_Article_9}. In principle, processing such data is \textbf{prohibited} unless a company meets one of the specific exceptions, listed in Art. 9(2) GDPR~\cite{EU_GDPR_Article_9}. In practice, for monitoring fairness in post-deployment, the only potentially relevant exception is the \textbf{explicit and freely given consent} of an individual, which requires that the individual has a genuine choice. According to the European Data Protection Board (an EU body responsible for GDPR application), consent is unlikely to be freely given in situations where there is a \textbf{power imbalance} between the individual and the company deciding on processing data (the data controller)~\cite{EDPB_consent,d2023market}. The employment context, the core setting of this paper, illustrates this imbalance, since job candidates may feel pressured to give their consent due to the potential negative consequences of refusing~\cite {EDPB_consent}. As a result, a hiring or recruiting company generally cannot rely directly on consent as a valid legal basis for processing sensitive data for fairness monitoring in recruitment contexts~\cite{EDPB_consent}. 

\revtext{To address these constraints, the protocol proposed in this paper introduces a \textbf{trusted third party} for collecting and processing personal data for fairness monitoring. In this setting, candidates provide consent to a trusted third party rather than the hiring company, which helps reduce the power imbalance. Several elements are required to support the legal validity of this approach: the third party is independent from the employer, the employer is not involved in data collection, consent is requested only after candidates receive their results (reducing pressure linked to potential negative consequences), and participation is fully voluntary and limited to fairness monitoring purposes. }       


\revtext{An important point to note is that the proposed approach also addresses a \textbf{limitation of the AI Act}. The Act provides an exception allowing the use of sensitive data for bias detection (Art. 10(5) AIA)~\cite{EU_AI_Act_Article_10}, but it is limited to the development phase (before the system is placed on the market), specifically to ensure data quality of training, validation, and testing \textbf{datasets} (Art. 10(2)(f)(g) AIA)~\cite{EU_AI_Act_Article_10}. Therefore, the legal basis of processing sensitive data for post-market monitoring fairness in the system's outputs, can rely only on the GDPR, as explained above. This further highlights the need for approaches such as the proposed protocol that enable fairness monitoring in a legally acceptable way. }

\noindent \textbf{Legal perspective on fairness.} 
In general, law does not rely on the concept of ‘fairness,’ widely used in computer science. Instead, EU law refers to the right to \textbf{non-discrimination}, one of the ‘fundamental rights’ recognized in the EU Charter of Fundamental Rights~\cite{EU_FUNDAMENTAL_RIGHTS}. Whether a practice constitutes illegal discrimination depends on a case-by-case assessment, considering all relevant circumstances; for example, whether a practice puts a person at a particular disadvantage and whether this disadvantage arises because of a protected characteristic. This means that even if an AI system appears ‘fair’ according to a particular fairness metric, a court may still find its outcomes discriminatory. On the other hand, a `particular disadvantage' may still be lawful if the practice is objectively justified, has a legitimate aim, and is proportionate. These elements cannot be assessed using a fairness metric alone. Nevertheless, multi-dimensional and multi-stage fairness metrics remain valuable for risk identification.

\subsection{Industry Requirements for Fairness Monitoring}\label{IndustryRequirements}

\textbf{Privacy and Data Requirements}: As noted in Section \ref{sec:legal-requirements}, the main industrial challenge arises from the legal tension between the need to detect and mitigate discrimination and the restrictions that the GDPR imposes on using sensitive data. Besides, collecting sensitive data can undermine candidate trust and damage a company's reputation if seen as intrusive or misused, so EU companies need compliant, transparent mechanisms, enabling fairness monitoring to be seen as a transparency-enhancing practice rather than a potential privacy threat. 

Building on the constraints identified in the challenges above, privacy and data requirements for fairness monitoring are defined by two main objectives: avoiding any direct access to sensitive data and enabling data collection without harming user trust. The industrial solution must therefore rely on privacy-preserving architectures grounded in legal certainty and limited strictly to fairness monitoring purposes. To meet these objectives, privacy-enhancing technologies and secure computation frameworks are required to enable collaborative or distributed analysis without centralizing individual-level data.


\noindent \textbf{Fairness Requirements}: Monitoring tools must support suitable fairness metrics across input, output, and outcome dimensions, including options for intersectional analysis. However, the interpretation of such metrics remains complex and often requires simplification or assisted visualization to remain usable in practice. Aggregation and comparison across company, job, and platform levels are essential, with a clear distinction between micro- and macro-averages to prevent loss of bias directionality. 

Additionally, systems should indicate minimum candidate counts to ensure that fairness metrics are interpretable and not excessively affected by statistical noise. Data quality in fairness monitoring remains a critical challenge, as low applicant trust and reluctance to share sensitive data may lead to sparse, biased, and incomplete samples, undermining the accuracy of fairness monitoring outcomes. Industrial users thus require recommendations on the minimum number of candidates per offer or per group to draw valid conclusions from the metrics, and expect reliability estimates for result interpretation. 


Establishing recommended thresholds or baselines is also essential. A metric will indicate potential discrimination if it falls below or above a certain recommended threshold, and defining these thresholds is necessary for relevant use and for triggering alerts from the monitors. However, for some metrics, it is challenging to establish a universal baseline group for comparison across diverse job titles, which necessitates both legal and empirical considerations.

Ultimately, systems should provide transparent criteria for defining when a candidate is considered qualified, as required for metrics such as Equal Opportunity. In large-scale recruitment settings, manually labeling candidate qualifications across multiple offers is unfeasible, making it necessary to rely on algorithmic matching scores to approximate qualifications. However, this approach still carries the risk of introducing bias into the definition itself.

\noindent \textbf{Usability Requirements}: Given the diversity of professional roles involved in fairness monitoring—ranging from data scientists and compliance officers to recruiters—the tools must facilitate the interpretation of metrics and monitoring outputs for both technical and non-technical users. Visualization is therefore a central requirement: results should be presented in dashboards that display fairness metrics, key performance indicators, and graphical plots in an accessible and comprehensible manner for all stakeholders.

Feedback received from companies shows that monitoring systems should also include the ability to track the historical evolution of metrics over time, capturing how fairness indicators evolve throughout the lifecycle of a job offer as candidate data accumulates. This temporal visibility enables organizations to detect trends, regressions, or improvements in fairness performance across recruitment cycles.

Usability also entails compliance functionality. The dashboards and generated reports should serve as verifiable documentation of compliance with the AI Act, allowing companies to use the same artefacts both for internal governance and for demonstrating conformity to regulators—thus avoiding redundant work and aligning fairness monitoring with existing compliance workflows. Beyond compliance, the possibility of using these monitoring tools as both an added incentive and a competitive distinction further reinforces their value.

\noindent \textbf{Feasibility Requirements}: From an engineering and deployment perspective, fairness monitoring solutions must be technically feasible for industry partners and align with existing production infrastructures. Consideration must be given to ensuring that all software components are stable, installable, and maintainable, avoiding dependencies on tools that lack sufficient technical maturity or cannot be deployed in corporate environments, while minimizing operational burden.



\section{Requirements-based System Design}

\begin{figure*}[htbp]
    \centering
    \includegraphics[width=1\textwidth]{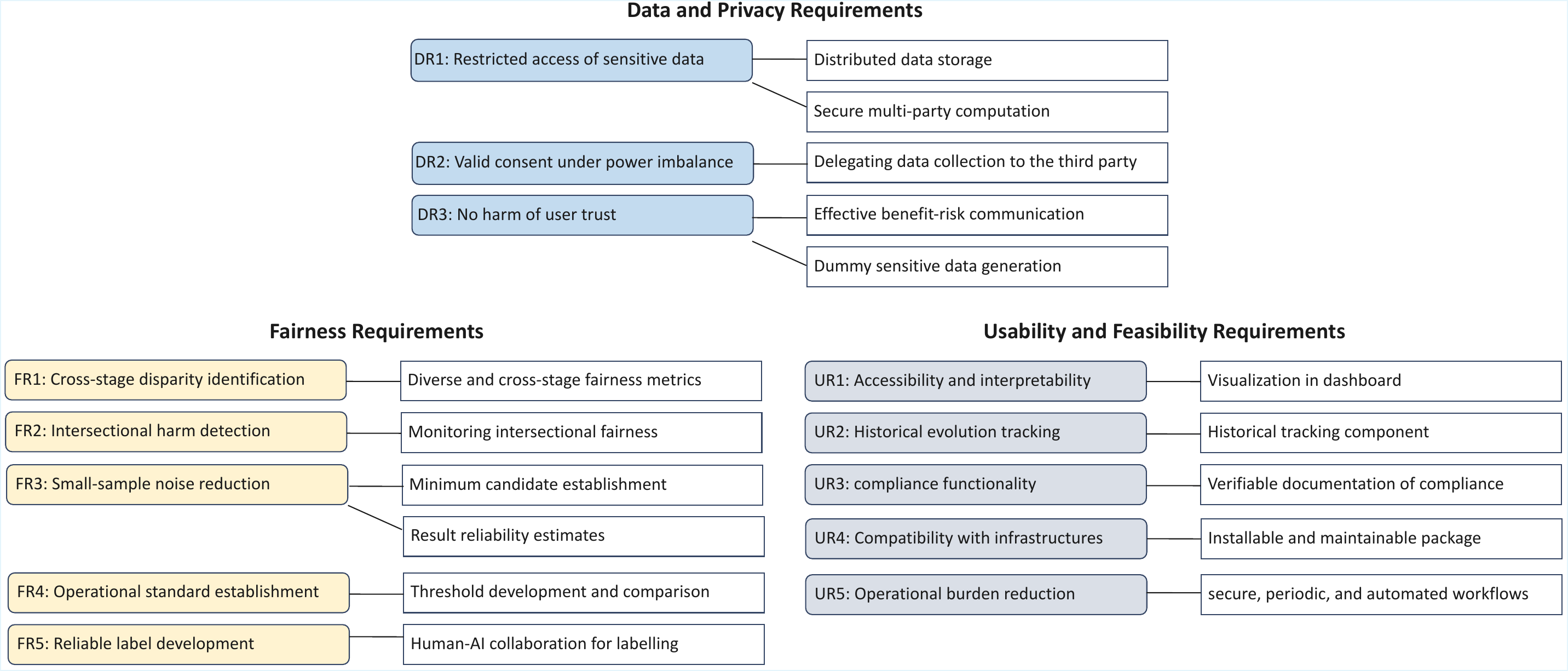}
    \caption{Design features of the MPC protocol for fairness monitoring based on design requirements.}
    \label{industry_to_design}
\end{figure*}

The system design features based on legal and industrial design requirements are illustrated in Figure \ref{industry_to_design}.

\subsection{Design Features for Data and Privacy}


Although the EU AI Act mandates post-market monitoring for high-risk AI systems~\cite{EU_AI_Act_Article_72}, the exception of processing special categories of personal data only applies to pre-deployment stages~\cite{EU_AI_Act_Article_10} (training, testing, or validation) rather than post-market monitoring. Access to sensitive data for post-market fairness monitoring remains restricted. As the first step, we propose \textbf{distributed data storage} combined with \textbf{secure multi-party computation} (MPC) to achieve fairness monitoring without allowing any party to access users' sensitive data. In particular, through secret sharing, users' sensitive data is distributed in secret shares between two parties: the hiring company as the deployer, and a trusted third party (TTP) that assists in encrypted data storage and fairness computation. Neither party could access users' raw sensitive data. Next, MPC allows for the joint computation of fairness metrics using inputs from the two parties without requiring any party to disclose the value of its inputs, thereby preserving the confidentiality of individual sensitive attributes in the computation process~\cite{goldreich1998secure,helminger2022multi}. The specific implementations are described in Section \ref{systemModel}.


As noted in legal requirements in Section \ref{sec:legal-requirements}, obtaining freely given consent in the hiring setting is challenging due to the power imbalance between individuals and companies. As distributed data storage and secure multi-party computation involve the assistance of the TTP (e.g., research centers or non-governmental non-profit organizations), \textbf{delegating data collection to TTP} emerges as a natural choice to mitigate concerns about data misuse and power imbalance. 
Furthermore, Section \ref{IndustryRequirements} suggests that collecting sensitive data during the hiring process inherently entails trust risks. Therefore, though the MPC protocol provides a promising approach for data protection, \textbf{effective benefit-risk communication} in data collection is essential for establishing meaningful consent~\cite{kacsmar2023comprehension,franzen2024communicating}. Moreover, to mitigate mistrust arising from concerns that data donation may influence hiring decisions, users' choices regarding whether to donate data should remain invisible to the hiring company. To this end, \textbf{dummy data generation}, which produces a secret share regardless of whether a user donates data, becomes a crucial step prior to distributing secret shares to the hiring company.

\subsection{Design Features for Fairness}

The legal and industrial requirements highlight several design features necessary for meaningful and reliable fairness monitoring. First, the monitoring system should provide \textbf{diverse and cross-stage fairness metrics} that encompass not only \emph{output fairness} in model predictions, but also \emph{input fairness} related to candidate pool diversity and \emph{outcome fairness} reflecting real-world hiring impacts. Besides, as harms often emerge at the overlap of multiple identities in real-world recruitment, supporting \textbf{the monitoring of intersectional fairness} is also crucial. Furthermore, real-world fairness monitoring is hindered by high statistical noise due to limited sample sizes, which undermines the reliability of the results. It not only necessitates the \textbf{establishment of minimum candidate counts} for generating meaningful monitoring results, but also highlights the need for \textbf{reliability estimates} (e.g., confidence interval and margin of error) along with fairness metrics. In addition, what constitutes ``sufficient fairness'' remains ill-defined and requires both legal and empirical consideration. \textbf{Defining clear fairness thresholds} for specific cases through collaboration among regulators, industry stakeholders, and civil-society organizations would further increase the monitoring practicality. Finally, some fairness metrics rely on labels (e.g., labels for ``qualified'' candidates for the computation of equal opportunity). Fully relying on manual labeling would be practically infeasible, while automated annotation assisted with external models could introduce new biases. A well-designed \textbf{human–AI collaboration pipeline for labeling} might offer a pragmatic balance between feasibility and reliability.

\subsection{Design Features for Usability and Feasibility}\label{DesignFeasibility}

The fairness monitoring system must be usable and feasible for real-world deployment. First, to afford interpretability for both technical and non-technical users (e.g., recruiters, AI engineers, external auditors), \textbf{visualizations in dashboards} that display fairness metrics and performance indicators would be beneficial. Besides, to track how metrics change over time throughout a job offer's lifecycle, monitoring systems should \textbf{capture the historical evolution of fairness indicators} as candidate data accumulates and evolves. Moreover, usability also requires built-in compliance support: the monitoring system should provide \textbf{verifiable documentation of compliance} that can serve both internal governance and external regulatory demonstrations of conformity. Finally, a fairness monitoring system needs to support compatibility with industrial infrastructures, and should thus be implemented as \textbf{installable and maintainable software packages}. To minimize operational burden in continuous monitoring, the systems should operate through \textbf{secure, periodic, and largely automated workflows} from data extraction and collection, encryption, fairness computation, to result generation and storage.

\section{The Framework of the MPC Fairness Monitoring Protocol}\label{systemModel}

\begin{figure}[htbp]
    \centering
    \includegraphics[width=0.8\textwidth]{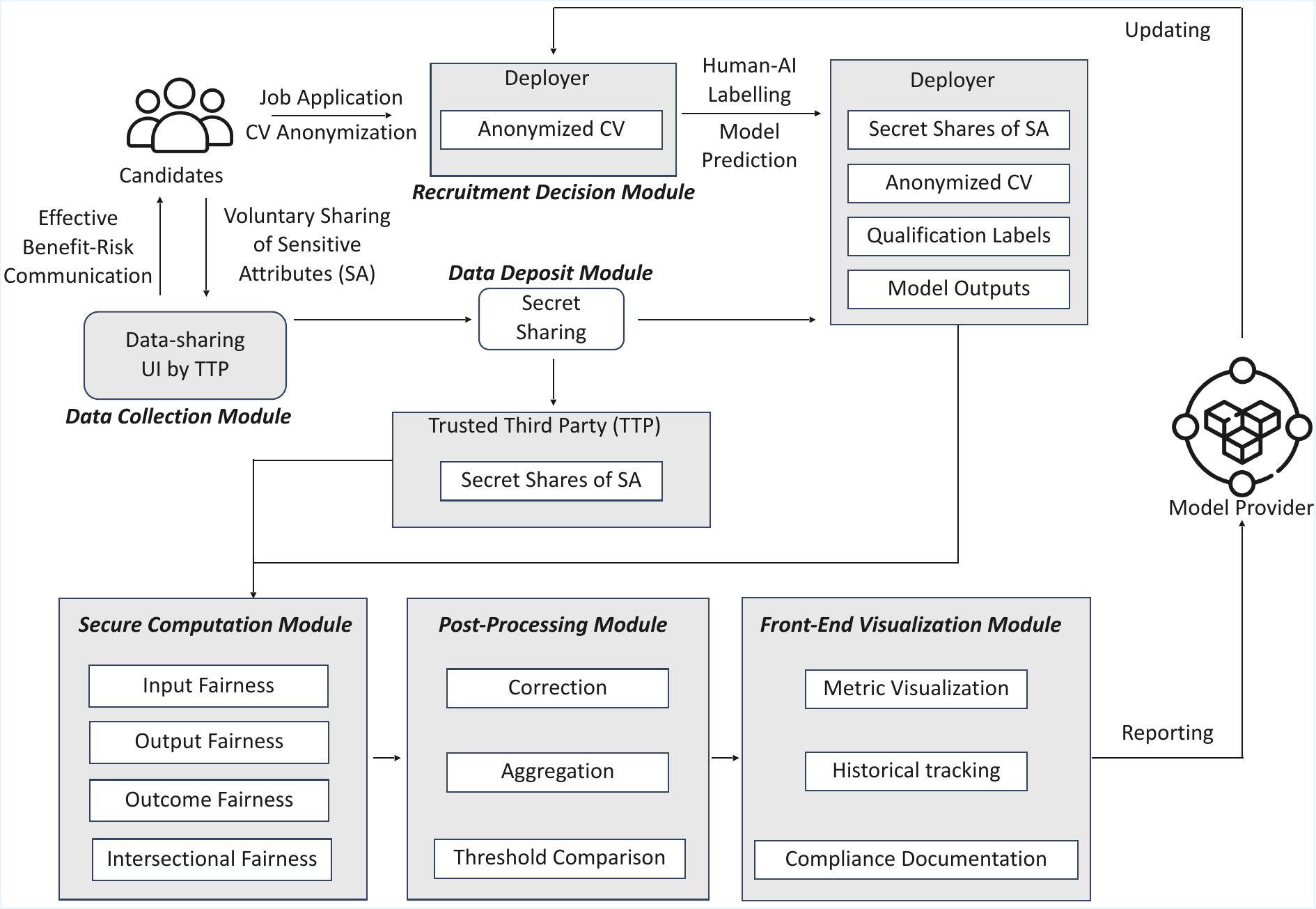}
    \caption{\revtext{The framework of the MPC protocol for fairness monitoring.}}
    \label{framework_overview}
\end{figure}

The framework of the MPC protocol for fairness monitoring is shown in Figure \ref{framework_overview}. In this framework, the \revtext{model} provider (typically the model developer) is responsible for establishing the MPC-based fairness monitoring system and ensuring regulatory compliance. The deployer, who operates the recruitment system, holds non-sensitive application data (e.g., anonymized CV), model outputs, and hiring outcomes. The trusted third party (TTP) collects user-provided sensitive data, generates secret shares, and manages sensitive data deposits. Then, the deployer and TTP participate in secure two-party computation to compute fairness metrics without directly accessing users' protected attributes. After post-processing, fairness monitoring results are presented in dashboard UI, and the deployer contributes to post-market monitoring by tracking fairness performance and reporting identified discrimination risks to the provider. The detailed system parties and data are shown in Appendix \ref{partiesData}.

\noindent \textbf{Problem overview.} Let $(X,Z,Y,G,D)$ be random variables \revtext{that vary across individuals}, where $X\in\X$ are non-sensitive features, $Z\in\Z$ system behavior variables, $Y\in\Y$ outcomes, $G\in\G$ protected group, and $D\in\{0,1\}$ indicates donation/availability of $G$. We observe $(X,Z,Y)$ for all individuals, but $G$ only when $D=1$.

We compute fairness functionals that can be written or closely approximated as linear functionals of group indicators:
\begin{equation}
T(P) \;=\; \sum_{g\in\G} \phi_g\!\left(\E\big[f(Z,Y)\cdot \I\{G=g\}\big]\right),
\label{eq:functional}
\end{equation}

where $f$ may encode exposure weights (ranking) or decision outcomes (classification), and $\phi_g(\cdot)$ aggregates to a metric such as demographic parity gap, equal opportunity gap, or exposure disparity.

\noindent \textbf{Recruitment decision module.} 
Operated by the deployer, this module maps non-sensitive application data to model outputs and final hiring outcomes. We consider a simplified decision-making model that incorporates both automated ranking and human assessment~\cite{fabris2025fairness}.
Given candidate features $X_i$, the automated hiring system produces a model score $S_i = s(X_i)$ and induces a corresponding ranking position $ R_i = \pi(S_i).$
Human evaluators then make a recruitment decision based on model outputs and contextual information, yielding the final outcome: $Y_i = d(S_i, R_i).$
The fairness monitoring protocol does not intervene in this process, but evaluates fairness based on the resulting distribution of $(Z_i, Y_i, G_i, D_i)$.

\noindent \textbf{Data collection module.}
Managed by a trusted third party (TTP), this module enables candidates to voluntarily share sensitive attributes. \revtext{By separating data collection from the employer, the protocol mitigates the power imbalance problem and supports the freely-given consent, thereby improving the prospects for lawful deployment.} It should ensure meaningfully informed consent with effective benefit-risk communication.


After submitting application materials, each candidate $i$ is redirected from the recruitment platform to a TTP-operated page with a pseudonymous linkage identifier $\ell_i$ that allows TTP to link data across parties while preventing the deployer from learning about the candidate's donation decision.

Upon redirection, the candidate may voluntarily provide sensitive attributes, \revtext{which are then mapped by the TTP to a numerical encoding $g_i$ of protected-group categories for secure computation}.
The TTP then defines 
$
D_i = 
\begin{cases}
1, & \text{if a sensitive attribute is provided}\\[1mm]
0, & \text{otherwise}
\end{cases}
$
and
$
G_i =
\begin{cases}
g_i, & \text{if } D_i = 1\\[1mm]
-1, & \text{if } D_i = 0
\end{cases}
$
where $-1$ denotes a dummy value that does not represent any valid sensitive category and will be excluded from subsequent fairness computations.

%
%

\revtext{To establish valid consent, data sharing decisions should not influence recruitment or its outcomes. With TTP-assisted data collection and dummy data generation, the deployer observes neither $G_i$ nor $D_i$, and thus cannot infer whether an individual candidate donated sensitive data. Additionally, collecting sensitive data after hiring may alleviate candidates’ concerns that disclosure could influence selection decisions; however, how hiring outcomes shape individuals’ willingness to share such data remains an open question for future research. Besides, the willingness to donate sensitive data may be influenced by the identity of the TTP (e.g., non-governmental organizations, research centers, specialized commercial companies). How to balance institutional governance, user trust, and technological capability when selecting a TTP remains a question that fosters further legal, social, and technical investigations~\cite{helminger2022multi,bogen2024navigating}. We discuss the technical debiasing options later in this section (post-processing module).}

\noindent \textbf{Data deposit module.}
After getting $G_i \in \mathcal{G}$, the TTP-operated data collection module immediately generates a random masking value $R_i$ and forms two secret shares: $
G_i^{(1)} = G_i + R_i, 
G_i^{(2)} = -\,R_i
$, so that
$
G_i^{(1)} + G_i^{(2)} = G_i.
$
The share $G_i^{(1)}$ is transmitted to the deployer, and the share $G_i^{(2)}$ is transmitted to TTP, each linked only to the candidate’s pseudonymous identifier $\ell_i$. 
After secret share generation and distribution, the raw sensitive attribute $G_i$ and masking value $R_i$ are immediately deleted.
These secret shares constitute the only persistent representation of sensitive attributes and serve as the joint input to the secure computation module.

\noindent \textbf{Secure computation module.}
The MPC module receives secret shares $\{G_i^{(1)}\}$ from the deployer and $\{G_i^{(2)}\}$ from the TTP. 
Within the secure computation environment, the two parties jointly form the masked reconstruction of the sensitive attribute
$
\widehat{G}_i \;=\; G_i^{(1)} + G_i^{(2)},
$
which is never revealed to either party.  
The reconstructed value $\widehat{G}_i$ exists only as an internal MPC variable and is never reconstructed in the clear.

Using $\widehat{G}_i$, the MPC computes the group-membership indicator for group $g$, i.e., $ \I(\widehat{G}_i = g) $, and uses it together with relevant visible variables from the deployer to compute fairness metrics. For example, to measure \textit{pool diversity} for input fairness, it is possible to use the group-membership indicator to compute the fraction of all $N$ candidates who belong to group $g$:
$
\mathrm{PD}(g)
=
\frac{
\sum_{i=1}^N \I(\widehat{G}_i = g)
}{N}
$ The fairness metrics should span different hiring stages (input fairness, output fairness, and outcome fairness) for a comprehensive evaluation. Several further examples across different stages can be found in Appendix \ref{metricsExample}. \revtext{We conducted a validation experiment using 
a synthetic dataset to verify the correctness of the MPC implementation, comparing fairness metrics between plaintext computation and MPC protocol-based computation. We present the details in Appendix \ref{correctnessCheck}. Across all experiments, the MPC outputs matched the plaintext results up to numerical precision (mean absolute error $\sim 10^{-9}$), confirming the functional validity of the protocol.}\footnote{\revtext{The differences between MPC-based fairness metrics and plaintext-based fairness metrics arise from secure computation constraints, where MPC uses secret sharing and fixed-precision arithmetic, requiring approximations for operations like division, logarithms, and normalization.}}

\textit{Intersectional Fairness.}
For multidimensional protected attributes, define $g=(g^{(1)},\ldots,g^{(m)})$ and compute
$
\I(\widehat{G}_i = g)
= 
\prod_{j=1}^m \I(\widehat{G}_i^{(j)} = g^{(j)}),
$
enabling all fairness metrics in Appendix \ref{metricsExample} to apply directly to intersectional groups. \revtext{Monitoring of intersectional metrics helps identify fairness issues that may affect individuals belonging to multiple protected categories and experience intersectional discrimination~\cite{wang2022towards,schlesinger2017intersectional}. Nonetheless, ensuring robust fairness monitoring as we deal with increasingly small intersectional subgroups remains a challenge~\cite{wang2022towards}.}

\textit{Feasibility of computing fairness metrics using MPC.}
The fairness metrics rely on operations of the form
$
a_i \cdot \I(\widehat{G}_i = g),
$
where $a_i$ is a non-sensitive scalar known to the deployer (e.g., $1$, $Z_i$, $Y_i$, $Q_i$, or $d(R_i)$), and $\I(\widehat{G}_i = g)$ is obtained through a secure equality test on secret shares. 
These computations require only standard MPC primitives: secure addition, secure multiplication by public scalars, and secure comparisons~\cite{zhao2019secure}.

Besides, the protocol also needs to filter out dummy values used for non-donors (e.g., $\widehat{G}_i = -1$). 
This filtering is implemented by an additional secure equality test
$
\I(\widehat{G}_i \neq -1),
$
and a Boolean conjunction represented as multiplication inside MPC, i.e.,
$
\I(\widehat{G}_i = g) \cdot \I(\widehat{G}_i \neq -1).
$
Such operations are lightweight and fully supported by standard MPC frameworks~\cite{zhao2019secure}.  
 

\noindent \textbf{Post-processing module.}
After the secure computation module outputs aggregated group-level fairness metrics, a post-processing layer refines these results to support interpretation, cross-context comparison, and actionable monitoring. This module does not access individual-level information and operates solely on group-level aggregates.

\textit{Reliability estimates.}
Because fairness metrics are computed from voluntarily donated samples, the robustness of each estimate depends on the sample size of contributing groups. \revtext{For example, for a proportion-based fairness metric $T(g)$, we can report an approximate $(1-\alpha)$ confidence interval using a normal approximation to the binomial distribution:
$
\widehat{T}(g) \pm z_{\alpha/2}\sqrt{\frac{\widehat{T}(g)(1-\widehat{T}(g))}{n_g}}, 
n_g = \sum_{i} \I(\widehat{G}_i = g)
$.
This approximation is valid when samples are independent and $n_g$ is sufficiently large. Such intervals allow auditors to distinguish genuine disparities from noise induced by scarce donation data. Besides, we suggest contextual considerations about sample sizes and potential sampling bias to select suitable intervals and facilitate accurate result interpretation. For example, alternatives such as Wilson or Agresti--Coull intervals may offer better small-sample performance.}


\textit{\revtext{Handling bias from voluntary data donation.}} \revtext{Because sensitive attributes are collected voluntarily, the donated dataset may not accurately represent the underlying candidate population, introducing selection bias if donation likelihood correlates with protected attributes or hiring outcomes (e.g., biases if members of a minority group are less willing to share data). To address this, the post-processing module should consider potential missing-data regimes (e.g., MCAR, MAR, MNAR~\cite{zhang2021assessing}) and apply corresponding debiasing strategies for missing-robust measurement (Appendix \ref{appendixMissingness}). }


\textit{Aggregation in organizational levels.}  
The module needs to support both micro- and macro-aggregation for job titles, departments, companies, or platforms:
$
T_{\mathrm{micro}} = \frac{\sum_{u} n_u T_u}{\sum_{u} n_u},
T_{\mathrm{macro}} = \frac{1}{U}\sum_{u} T_u,
$
where $u$ indexes organizational units and $n_u$ denotes their sample sizes.  
Micro-averages weigh large units more heavily, while macro-averages treat all units equally. Reporting both micro- and macro-aggregation across different organizational levels prevents masking bias directionality, e.g., a minority group disadvantaged in several small units but obscured by a large neutral unit.

\textit{Thresholding and alerting.}  
Interpreting disparities requires recommended thresholds indicating when a metric suggests potential discrimination.  
A group disparity $\Delta(g)$ may trigger a warning when
$
|\Delta(g)| \ge \tau,
$
where $\tau$ is a legally or empirically informed tolerance parameter.  
Thresholds must account for job-specific context, since no universal baseline group exists for all metrics or recruitment settings.  
The post-processing module therefore allows configurable thresholds, alert rules, and documentation of the underlying assumptions for their use.

This post-processing module transforms raw fairness metrics into interpretable, comparable, and actionable monitoring signals, without interacting with individual-level data or compromising the privacy guarantees of the protocol.

\noindent\textbf{Front-end visualization module.}
To support usability in real-world deployment, the monitoring system should provide a front-end interface that enables intuitive interpretation of fairness results for both technical and non-technical stakeholders. As detailed in Sec. \ref{DesignFeasibility}, the front-end interface should include at least: (1) a dashboard that visualizes group-level fairness metrics, threshold comparisons, reliability indicators, and performance summaries; (2) temporal visualizations to show the evolution of fairness metrics as candidate data accumulates over a job offer's lifecycle; (3) verifiable, exportable documentation of monitoring results to support internal governance workflows and demonstrate regulatory compliance.

\noindent \textbf{\revtext{Adversarial model and privacy guarantees.}} \revtext{We assume a semi-honest (honest-but-curious) adversary model, where parties follow the protocol but may attempt to infer additional information from the data they observe. The deployer holds the non-sensitive hiring data $(X, Z, Y)$ and the secret shares $G^{(1)}$, while the trusted third party (TTP) holds the secret shares $G^{(2)}$ derived from voluntarily donated sensitive attributes. We assume no collusion between the two parties. This assumption aligns with real-world deployment scenarios where the deployer and the TTP represent independent organizations with distinct governance and incentives (e.g., a company and an external auditing service), reducing the risk of collusion while allowing secure MPC-based monitoring.}

\revtext{Under this model, the protocol provides input privacy following standard semi-honest secure multi-party computation (MPC) guarantees: each party learns nothing about the other party's inputs beyond what is implied by its own input and the protocol output. In particular, neither party can reconstruct individual-level sensitive attributes, and the protocol reveals only the aggregated fairness metrics. Besides, to mitigate side-channel leakage, dummy shares are created for non-donors to keep the computation independent of sensitive-data donation, and the donation indicator $D$ is hidden during MPC execution. }



\section{Implementation: Industrial Dashboard}\label{IndustrialDashboard}

The industry members of the research team implemented the proposed post-market fairness monitoring system as a \textbf{demonstrator} integrated with their internal production data pipelines, and visualized through the industrial dashboard shown in the Fig.~\ref{fig:dashboard} in Appendix \ref{appendixC}. The implementation aimed to validate the technical feasibility of the protocol under real-world constraints, while respecting data protection limitations and existing organizational workflows. 

\revtext{The industrial deployment was executed on a distributed Databricks Spark cluster with 1--12 workers (ranging from 16--192 GB memory and 4--48 CPU cores). The evaluation covered 1,778,532 applications from 687,638 unique candidates across 8,429 job offers and 558 distinct job titles ranging from January 2024 to June 2024, without discarding any data. Completion of the monitoring pipeline took approximately 46 minutes. This large-scale deployment demonstrates the feasibility of integrating MPC-based fairness monitoring into production-grade infrastructures while maintaining acceptable computation times for periodic monitoring workflows.}

\noindent \textbf{Data pipeline and scope.} The first implementation step consisted of identifying the data sources available in production and designing dedicated pipelines to construct the datasets required for fairness monitoring. A periodic extraction process retrieves updated job-offer data from the Production Database (DB), focusing on offers created or updated in the latest interval to exclude inactive offers. In practice, the pipeline covers \revtext{about} one year of historical data.

For each offer, the extracted data includes: (i) offer-level metadata such as job title classification and company identifier; and (ii) candidate screening information, including applications and non-sensitive variables denoting progression in the hiring pipeline (e.g., relevance scores or indicators of moving forward to subsequent stages). These variables correspond to the non-sensitive inputs and outcomes described in the system model and serve as the basis for computing exposure- and outcome-based fairness metrics. 

\noindent \textbf{Use of \revtext{personal} attributes.} The implemented system uses \textbf{Age} and \textbf{Gender} that candidates may voluntarily provide as part of standard platform usage, which constitute personal data but do not fall under the special categories of personal data under Article~9 GDPR~\cite{EU_GDPR_Article_9}. 
\revtext{This choice reflects practical and legal feasibility considerations: using attributes already available in the production environment enables validation under realistic conditions without introducing additional data collection or user-facing changes. In contrast, incorporating special-category data would require more complex GDPR compliance measures and additional data collection, likely limiting the evaluation to a controlled pilot rather than a production setting. The technical feasibility, correctness, and privacy guarantees of the MPC protocol are attribute-agnostic. Regarding the age attribute}, because most fairness metrics operate over categorical group variables, the continuous age attribute was discretized into three categories based on the empirical age distribution of candidates on the platform: ``$<27$'', ``$27$--$37$'', and ``$>37$'' years. 





\noindent \textbf{Simulated TTP and secure computation.} The TTP role was simulated internally for demonstrative purposes. This simulation was implemented by generating and storing two separate and independent encrypted datasets: one associated with the deployer and one associated with the simulated TTP. Each dataset contains only one additive secret share of the sensitive attributes, ensuring that neither party alone can reconstruct individual-level protected attributes.

Fairness monitoring functions are implemented using the open source MPC protocol in Section \ref{systemModel}, which jointly processes the secret shares to compute aggregated fairness metrics—pool diversity, group exposure, demographic parity, equal opportunity, and their intersectional variants—without revealing individual sensitive attributes or donation decisions. Only non-sensitive, group-level results are stored in the production database.

\noindent \textbf{Dashboard and visualization.} The final step translates the protocol outputs into an interpretable visualization using a Tableau\footnote{https://www.tableau.com/}-based dashboard (Fig.~\ref{fig:dashboard}). The dashboard supports daily monitoring and comparative analysis across four aggregation levels: individual job offers, job title classifications, companies, and overall. It presents fairness indicators through a combination of key performance indicators (KPIs), pie charts for comparative analysis, histograms \revtext{of} metric distributions, and bar plots contrasting observed metric deviation and actual values. It also includes a \textit{historical evolution dashboard}, \revtext{tracking} how fairness metrics evolve over the \revtext{offer} lifecycle as candidate data accumulates. In this way, the dashboard operationalizes the outputs of the privacy-preserving protocol into actionable insights for fairness monitoring and governance.


\section{Lessons and Practical Implications}\label{lessons}



\noindent \textbf{Legal learnings: compliance challenges, open questions, and the insights for the Omnibus proposal.} \revtext{The proposed monitoring protocol is aligned with GDPR requirements through several design elements. The protocol relies on a trusted third party for data collection and processing, which is structurally independent and separated from the employer, reducing the power imbalance present in employment contexts. In addition, candidates provide consent only after receiving the hiring outcome, which ensures that disclosure of sensitive data does not influence hiring decisions. Participation in data donation is fully voluntary and limited to fairness monitoring purposes. Taken together, these elements are intended to meet the conditions for valid consent.
At the same time, there are a number of open questions that are primarily policy-oriented. The effective implementation of the protocol would benefit from further institutional and regulatory guidance, for example on who can qualify as a trusted third party, how such entities should be selected or governed, and how frequently monitoring should occur. As such, the protocol should be understood as a procedural proposal that could be supported and refined through EU-level guidelines, and that deserves broader legal and policy debate.}

\revtext{These considerations become even more relevant in light of recent regulatory proposals under the EU Digital Omnibus Package, aimed at aligning EU digital laws~\cite{OmnibusProposal}. The AI Omnibus introduces a new exception that allows deployers (not only providers) to use sensitive data to detect and correct bias in AI systems, also non-high risk. However, this exception is still limited to training, validation, and testing datasets, and does not extend to monitoring system outputs in post-market monitoring. Therefore, even if the Omnibus is adopted (its final scope remains the subject of negotiation), the proposed protocol can serve as an important compliance safeguard as practical details remain under-specified.}


\noindent \textbf{Organizational and governance learnings.} From an industry perspective, the development and deployment of the fairness monitoring demonstrator yielded several organizational and technical insights; Table~\ref{tab:executive-lessons} in Appendix~\ref{appendixC} provides an overview of the key lessons. First, the demonstrator contributed to strengthening the company’s \textbf{AI governance and compliance foundation}. The project, with its network of experts and profound knowledge,  fostered a shared understanding of responsible AI practices across teams and helped clarify the interplay between the AI Act, GDPR, and non-discrimination law. In this sense, the initiative served as a concrete preparatory step toward future AI Act compliance by translating abstract regulatory requirements into operational processes and actionable measures.

Second, the work highlighted the importance of a \textbf{multidisciplinary approach to responsibility}. One of the main complexities identified was the gap between different perspectives: internally, across roles involved in AI projects (e.g., product, data, engineering, legal, privacy, security, UX, and communications), and externally, between academic research, legal interpretation, and business realities. The demonstrator acted as a catalyst for formalizing early and continuous involvement of these roles in AI projects, reinforcing the need for a 360-degree governance approach from the design stage \cite{iapp_ai_governance_ecosystem_2025}. At the same time, the experience confirmed that sustained collaboration between academia, industry, and the public sector is essential to align technical solutions with EU regulatory principles.

Third, the project surfaced a recurring \textbf{data centrality and sensitive data dilemma}. Fairness monitoring depends critically on data quality, representativeness, traceability, and justification. However, the need for protected attributes to detect and mitigate discrimination directly conflicts with legal and ethical constraints on processing sensitive data in hiring contexts. To avoid implementation paralysis, the demonstrator deliberately relied on voluntarily shared age and gender attributes, accepting reduced coverage in exchange for legal certainty and deployability.

Finally, the demonstrator reinforced that \textbf{trustworthiness extends beyond algorithms}. Trust in AI systems is shaped not only by models and metrics, but also by interface design, communication, and transparency mechanisms. User feedback played a key role in shaping the dashboard as an interface for transparency and explainability.

Responsibility involves not only assessing \textbf{what} technology is used and its implications, but also \textbf{why} that specific technology is chosen in the first place. Importantly, design choices were treated as non-neutral decisions \cite{winner1980artifacts} with technical, social, and ethical implications, highlighting the need to explicitly document and justify these choices to ensure traceability and auditability throughout the system lifecycle.

\noindent \textbf{Technical and implementation learnings.} 
From a technical standpoint, several additional lessons emerged. The \textbf{effort and development cost} of the demonstrator required approximately five months of focused effort by a data scientist with expertise in Python, Extract-Transform-Load (ETL) processes, and dashboarding, supported by other data professionals and a product manager. Most of the effort was devoted to (i) designing and maintaining robust ETL pipelines that correctly joined multiple production data sources; (ii) precomputing fairness metrics at different aggregation levels to avoid performance bottlenecks in the dashboard; and (iii) iterating on dashboard design and visualizations to improve usability and practical value.

The demonstrator confirmed the \textbf{feasibility of fairness monitoring dashboards} based on real production data. Simulating secret sharing and secure multi-party computation showed that these privacy-preserving techniques do not inherently limit the computation of fairness metrics or their integration into monitoring workflows. Crucially, the dashboards were developed and reviewed in parallel with the legal team, ensuring that metric selection, interpretation, and presentation were aligned with legal requirements and capable of withstanding regulatory scrutiny.

At the same time, the implementation revealed significant \textbf{complexity in interpretation and decision-making}. The main challenge was not implementing fairness metrics, but ensuring that diverse user groups—such as AI developers, product teams, sales teams, and clients— could interpret the dashboard results correctly and draw appropriate conclusions. \revtext{This complexity is compounded by the presence of multiple KPIs and comparison dimensions, as fairness metrics are analyzed not only at the level of individual job offers, but also in relation to job title and company-level aggregates.} The inclusion of multiple groups and metrics, particularly intersectional metrics, \revtext{further increases interpretative difficulty.} This required additional guidance on minimum sample sizes per offer or group, and careful consideration of thresholds. 

The difficulty of defining legally and contextually appropriate thresholds limited the immediate use of the system as an automated alerting mechanism. \revtext{In particular, fairness thresholds are inherently context-specific~\cite{corbett2023measure} and cannot be reduced to fixed or universal benchmarks. Accordingly, the framework supports configurable and explicitly documented thresholds (see Sec.~\ref{systemModel}), which are intended to function as indicators for \emph{risk flagging} rather than as definitive assessments of discrimination. This distinction is critical in light of EU non-discrimination law, where the evaluation of discriminatory effects is inherently contextual and case-specific, taking into account broader social, legal, and factual circumstances. Unlike heuristic rules such as the U.S.\ “80\% rule,” which rely on predefined numerical cut-offs, European legal frameworks do not prescribe uniform quantitative thresholds~\cite{wachter2021fairness,weerts2023algorithmic}.} All of these technical challenges had to be framed within legal boundaries: interpretations, thresholds, and decisions needed to remain aligned with applicable legal requirements.




\section{\revtext{Limitations}}

\revtext{Our study develops an infrastructural and socio-technical fairness monitoring framework under a concrete regulatory setting, rather than providing a universal specification for practical monitoring operations. We validated the feasibility of MPC monitoring infrastructure,  implementing core modules in an industrial environment, from recruitment decision module, data deposit module, secure computation module, to front-end visualization module. However, certain design features, such as TTP-assisted data collection, benefit-risk communication, minimum donation size requirement, and human-AI collaboration for labeling, necessarily require contextual and legal determination prior to deployment. In future applications, these design elements should be operationalized based on real-world contexts and practical needs.}


\section{Conclusion}

Multi-party computation (MPC) enables secure post-market fairness monitoring without exposing sensitive attributes; however, its practical deployment in real-world hiring processes, subject to legal, industrial, and usability constraints, remains unclear. To address this gap, this work developed MPC-based fairness monitoring protocols grounded in real-world requirements. We adopted a co-design approach that integrated technical, legal, and industrial expertise in algorithmic hiring. We identified key legal and industrial design requirements for MPC fairness monitoring, including data and privacy, fairness, usability, and feasibility considerations. Based on these requirements, we designed an end-to-end MPC fairness monitoring framework covering data collection, encryption, fairness computation, and front-end presentation. We empirically validated the framework through a fairness monitoring dashboard deployed in a large-scale, real-world industrial setting. Based on the study, we discussed the legal and industrial implications of deploying MPC-based post-market fairness monitoring for algorithmic hiring.

\section{Generative AI Usage Statement}

The authors used LLM (ChatGPT 5.1) solely for language polishing, including grammar correction and word refinement. No original content or arguments were generated by the tool.

\begin{acks}
This work has been supported by the FINDHR project, Horizon Europe grant agreement ID: 101070212. We thank all the people involved in the FINDHR consortium for productive discussions. We also particularly thank Deepak Garg, Carmela Troncoso, Peter Schwabe, Gulio Malavolta, Carlos Castillo, Alessandro Fabris, Didac Fortuny, David Graus, Antonio Mastropietro, Volodymyr Medentsiy, Salvatore Ruggieri, and Clara Rus for their insightful suggestions during the development of this study. 
\end{acks}

\bibliographystyle{plain}
\bibliography{sample-base}

\appendix

\section{System Parties and Data}\label{partiesData}

\subsection{Parties}

\begin{itemize}

    \item \textbf{\revtext{Model} Provider}: The \revtext{model} provider is responsible for establishing and operating the MPC-based fairness monitoring system and ensuring ongoing regulatory compliance. It updates the hiring model as necessary based on fairness monitoring reports provided by the deployer. It provides necessary technical support to deployers and the TTP to ensure the feasibility and interoperability of the MPC framework, without accessing raw individual-level data.
    
    \item \textbf{Deployer}: The entity operating the recruitment system, typically the employer. It holds non-sensitive application data (e.g., anonymized CV), model outputs, and hiring outcomes, and participates in secure two-party computation without directly accessing users' protected attributes. The deployer contributes to post-market monitoring by tracking fairness performance and reporting identified discrimination risks to the provider.
    
    \item \textbf{Trusted Third Party (TTP)}: An independent entity responsible for collecting user-provided sensitive data, generating secret shares, and managing sensitive data deposits. The TTP cooperates with the deployer through secure two-party computation to compute fairness metrics without disclosing individual sensitive attributes.

    \item \textbf{Candidates}: Individuals applying for job positions. They provide application data to the hiring company, acting as the deployer of the hiring model, and may voluntarily share sensitive attributes for fairness monitoring purposes. The recruitment process and hiring outcomes shall remain independent of any disclosure of sensitive attributes for monitoring.

    \item \textbf{Data Donors}: A subset of job candidates who provide explicit and voluntary consent to disclose their sensitive attributes (e.g., gender, age, ethnicity) for fairness monitoring. The decision to share data should not affect the recruitment process or its outcomes.

    \item \textbf{Human Recruiters}: Human decision-makers responsible for reviewing ranked candidates and determining qualification assessments or final hiring decisions. Their evaluative behavior constitutes an integral component of the sociotechnical hiring process and should be taken into account in the design, evaluation, and interpretation of fairness metrics.

    \item \textbf{Regulators/fairness auditors}: The supervisory actors responsible for overseeing compliance with applicable legal and ethical requirements in the operation of fairness monitoring procedures.
    
\end{itemize}

\subsection{Data}

\begin{itemize}

  \item \textbf{Non-sensitive application data.} Non-sensitive application data of candidates collected during the recruitment process. It includes data types such as education, skills, and employment history, which reflect job-related qualifications and serve as the basis for evaluation and selection. The non-sensitive application data is held by the deployer.
  
  \item \textbf{Sensitive attributes.} Sensitive attributes such as gender, age, ethnicity, or disability status that are legally protected and used exclusively for fairness assessment. Sensitive attributes are collected by TTP via voluntary donation, and must be stored exclusively as secret shares. 
  

  \item \textbf{Linkage identifiers.} Pseudonymous identifiers exchanged between the deployer and TTP to match job application records with sensitive-attribute deposits without revealing user identity.

  \item \textbf{Two-party secret shares.} Complementary shares generated through additive secret sharing from user-provided sensitive attributes. For each sensitive attribute, one share is stored by the deployer, the other by TTP. Neither party alone can reconstruct sensitive attributes based on the secret shares, but they can be jointly used for secure MPC fairness computations.

  \item \textbf{System behavior variables.} Contextual variables that characterize how the sociotechnical hiring system operates. These variables are not part of the candidate's application data and are not model predictions, but rather derived from human decision-making processes or empirically constructed behavioral models. Examples include (1) attention weights used for \textit{group exposure}, estimated from empirical observations of how recruiters inspect ranked candidates; (2) shortlist sizes used for \textit{top-k fairness}, reflecting organizational recruitment policies or constraints; (3) qualification assessments used for \textit{equal opportunity}, indicating whether a candidate meets job-specific criteria based on human evaluation or structured assessment processes. These system behavior variables provide essential information for modeling real-world recruitment dynamics and enable fairness metrics to reflect not only algorithmic outputs but also the broader hiring practices within which the automated system operates.

  \item \textbf{Model outputs.} The predictions of the automated hiring system, e.g., the rankings or scores of candidates for specific job titles.

  \item \textbf{Hiring outcomes.} The decisions of the hiring system, such as whether a candidate was shortlisted, interviewed, or hired.
  
  \item \textbf{Donation indicator.} Whether a candidate donates the sensitive attributes for fairness monitoring. The deployer must not be able to infer the donation indicator from protocol outputs or side channels.

  
\end{itemize}

\section{An Example Set of Fairness Metrics that Span Different Hiring Stages}\label{metricsExample}

\begin{enumerate}
    \item \textbf{Input fairness — Pool diversity.}
    Pool diversity measures how well each protected group is represented in the overall candidate pool. 
    For a group $g$, it is defined as:
    \[
    \mathrm{PD}(g)
    =
    \frac{
    \sum_{i=1}^N \I(\widehat{G}_i = g)
    }{N},
    \]
    that is, the fraction of all $N$ candidates who belong to $g$.
    
    \vspace{0.8em}
    \item \textbf{Output fairness — Group exposure.}
    Group exposure captures how much recruiter attention is received by members of a given group, taking into account that higher-ranked candidates are more likely to be inspected.  
    Let $R_i$ denote the rank of candidate $i$ in a given list and let $w(\cdot)$ be a position-based attention model (e.g., empirically estimated inspection frequency or a theoretical decay function).  
    We define the exposure weight
    \[
    Z_i \;=\; w(R_i),
    \]
    which is non-sensitive and known to the deployer.  
    The exposure of group $g$ is then
    \[
    \mathrm{GE}(g)
    =
    \frac{
    \sum_{i=1}^N Z_i \, \I(\widehat{G}_i = g)
    }{
    \sum_{i=1}^N Z_i
    }.
    \]
    that is, all the attention assigned to candidates in group $g$ divided by the total attention across the pool.
    
    \vspace{0.8em}
    \item \textbf{Output fairness — Top-$k$ fairness.}
    Top-$k$ fairness focuses on the most influential part of the ranking, i.e., candidates in the first $k$ positions. 
    Let $k$ be a user-specified cut-off (e.g., the empirical shortlist size) and $\mathcal{T}_k = \{ i : R_i \le k \}$ be the set of candidates in the top-$k$.
    
    \textit{(a) Skew@k.}
    Skew@k compares the proportion of group $g$ in the top-$k$ with its proportion in the full pool:
    \[
    \mathrm{Skew@k}(g)
    =
    \frac{
    \sum_{i \in \mathcal{T}_k} \I(\widehat{G}_i = g)
    }{k}
    -
    \frac{
    \sum_{i=1}^N \I(\widehat{G}_i = g)
    }{N}.
    \]
    A negative value indicates that group $g$ is underrepresented among the top-$k$ candidates relative to its overall presence in the pool, whereas a positive value indicates overrepresentation. Skew@k is appropriate when only inclusion in the top-k pool matters (e.g., shortlisted for interviews) while the ranking does not influence subsequent decisions.
    
    \textit{(b) Discounted Representation Difference (DRD@k).}
    While Skew@k looks only at membership in the top-$k$, DRD@k also accounts for the exact rank positions within the top-$k$ by applying a discount function.  
    Let $d(\cdot)$ be a monotonically decreasing function of the rank (e.g., $d(R) = 1/\log_2(R+1)$).  
    The discounted representation difference for group $g$ is
    \[
    \mathrm{DRD@k}(g)
    =
    \sum_{i \in \mathcal{T}_k}
    d(R_i)\,\I(\widehat{G}_i = g)
    -
    \sum_{i \in \mathcal{T}_k}
    d(R_i)\,\I(\widehat{G}_i \neq g).
    \]
    Here, higher-ranked positions contribute more through larger $d(R_i)$, so disparities near the very top of the ranking are weighted more strongly. DRD@k is most suitable when higher-ranked positions carry more impact (e.g., integrating CV rankings and subsequent interview-based evaluation to make hiring decisions).
    
    \vspace{0.8em}
    \item \textbf{Outcome fairness — Demographic parity.}
    Demographic parity evaluates whether different groups receive positive outcomes (e.g., being shortlisted, interviewed, or hired) at similar rates, regardless of underlying qualification.  
    Let $Y_i \in \{0,1\}$ denote a binary positive outcome for candidate $i$.  
    The demographic parity rate for group $g$ is
    \[
    \mathrm{DP}(g)
    =
    \frac{
    \sum_{i=1}^N Y_i \,\I(\widehat{G}_i = g)
    }{
    \sum_{i=1}^N \I(\widehat{G}_i = g)
    }.
    \]
    A large difference in $\mathrm{DP}(g)$ across groups indicates that some groups systematically obtain positive outcomes at higher or lower rates than others.
    
    \vspace{0.8em}
    \item \textbf{Outcome fairness — Equal opportunity.}
    Equal opportunity refines outcome fairness by conditioning on candidates who are deemed qualified.  
    Let $Q_i \in \{0,1\}$ denote whether candidate $i$ is qualified according to a pre-defined human or human–AI assessment process.
    Equal opportunity for group $g$ is defined as
    \[
    \mathrm{EO}(g)
    =
    \frac{
    \sum_{i=1}^N Y_i\,Q_i\,\I(\widehat{G}_i = g)
    }{
    \sum_{i=1}^N Q_i\,\I(\widehat{G}_i = g)
    }.
    \]
    This metric compares the probability of a positive outcome across groups among candidates who are all considered qualified.  
    If $\mathrm{EO}(g)$ is lower for a particular group, it indicates that qualified candidates from that group receive positive outcomes less often than equally qualified candidates from other groups.
\end{enumerate}

\section{\revtext{Correctness Evaluation of MPC Protocol Compared to Plaintext Baseline}}\label{correctnessCheck}

\begin{figure*}[htbp]
    \centering
\includegraphics[width=0.8\textwidth]{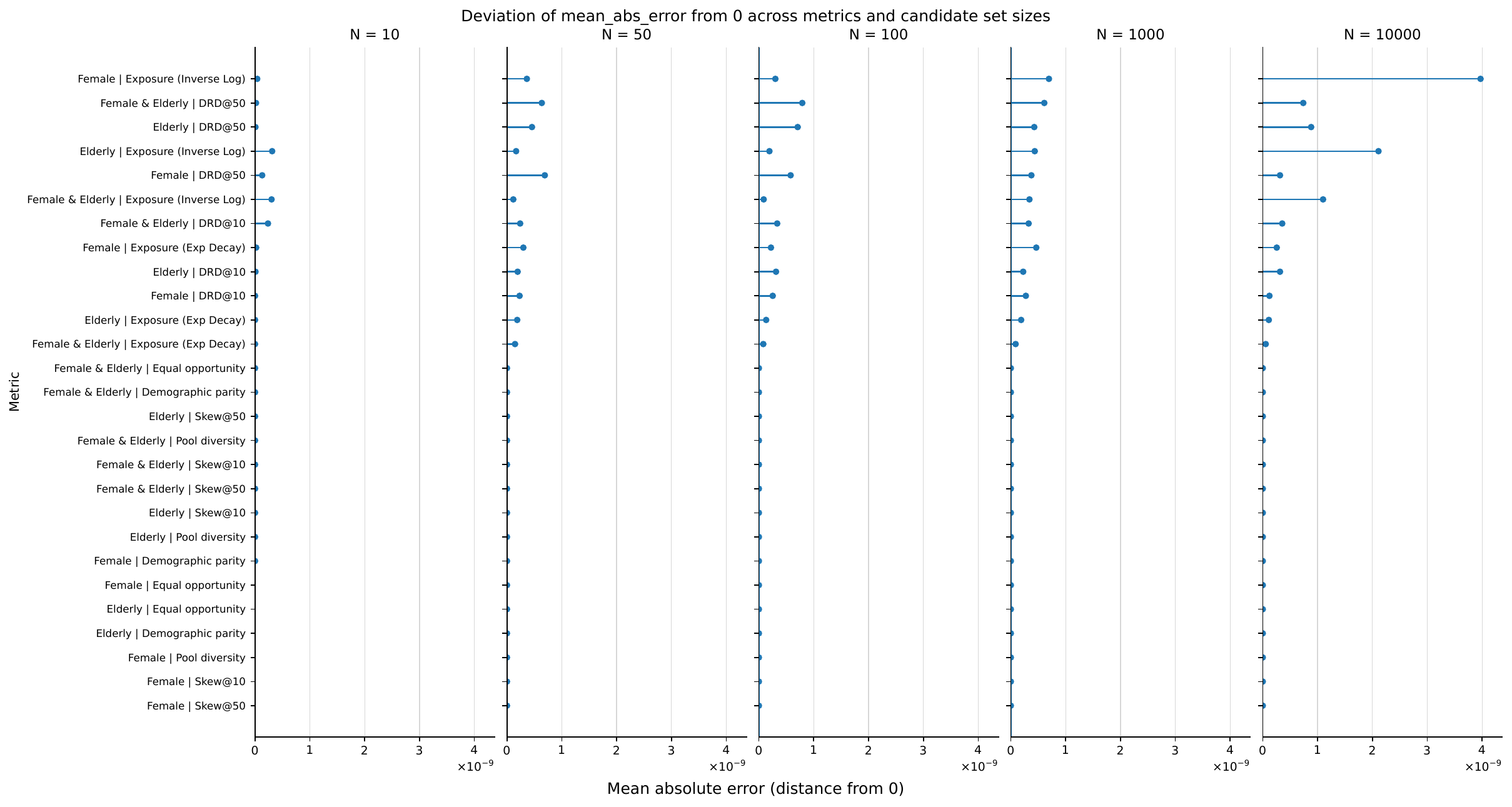}
    \caption{\revtext{Correctness evaluation on synthetic data: MPC-based fairness metrics vs. plaintext fairness metrics. We evaluate multiple fairness metrics across different protected groups (female, elderly, and their intersection) in different pool sizes to verify robustness of MPC computations across diverse fairness definitions, group structures, and pool sizes. The mean absolute error is on the order of $10^{-9}$ across all settings, indicating near-perfect agreement with the plaintext baseline.}}
    \label{correctnessCheckFig}
\end{figure*}

\revtext{To evaluate the correctness of the MPC Protocol, we compared MPC-based fairness metrics with plaintext fairness metrics on synthetic datasets, as shown in Figure \ref{correctnessCheckFig}. The synthetic datasets simulate a biased hiring pipeline with controlled statistical properties. For each trial, $N \in \{10, 50, 100, 1000, 10000\}$ candidates are generated with $P(\text{female}) = 0.43$ and $P(\text{elderly}) = 0.13$. Features include $\text{years of experience} \sim \mathcal{N}(6.5, 2.2)$ to simulate mid-career experience distribution, $\text{education} \in \{0,1,2\}$ representing low, medium, high education level, and $\text{test score} \sim \mathcal{N}(0,1)$. A latent quality score is computed as a noisy linear combination of features with explicit penalty terms $(-0.18\ \text{female},\ -0.22\ \text{elderly},\ -0.10\ \text{intersection})$ subtracted from the quality score. This latent score is transformed via a sigmoid function to obtain $P(\text{qualified})$. Candidates are ranked by a noisy score derived from this latent value, and acceptance depends on qualification and exceeding a score threshold (top $28\%$), again with demographic penalties. This parameterization introduces known disparities in both ranking and selection, enabling controlled evaluation of fairness metrics.}


\revtext{The evaluation has covered varying pool sizes ($N=10,50,100,1000,10000$) and multiple fairness metrics. We compared the differences between MPC-based fairness metrics and plaintext fairness metrics in 5 trials for each setting to mitigate stochastic variability. \textbf{The mean absolute error is on the order of $10^{-9}$ across all metrics and candidate sizes, indicating that the MPC implementation achieves near-perfect correctness relative to the plaintext baseline.} Differences between MPC-based and plaintext fairness metrics arise from the constraints of secure computation: in MPC, values are represented as secret shares and computed using fixed-precision secure types, which require approximations for operations such as division, logarithms, and normalization.}

\section{\revtext{Missingness Mechanisms of Voluntary Data Collection and Solutions}}\label{appendixMissingness}


\revtext{Under the voluntary basis of data collection, the dataset for fairness monitoring may not fully represent the candidate population. In this section, we discuss possible missingness mechanisms and potential solutions.}

\begin{itemize}
    \item \revtext{\textbf{MCAR (Missing Completely at Random):} Donation is unrelated to either observed features or sensitive attributes, which is nearly impossible. In this case, fairness metrics based on donated samples remain approximately unbiased, and standard uncertainty estimates could reflect sampling variability.}

    \item \revtext{\textbf{MAR (Missing at Random):} Donation may depend on observable non-sensitive attributes $X$ (e.g., applicants with higher education levels may be more likely to donate) but not on the sensitive attribute itself once $X$ is controlled for. In this regime, bias may be partially mitigated using auxiliary information from $X$, for example, by predicting group membership from observable attributes or by reweighting metrics using estimated donation propensities.}

    \item \revtext{\textbf{MNAR (Missing Not At Random):} Donation depends directly on the sensitive attribute (e.g., members of a minority group may be less likely to disclose it). Despite its high possibility in practice, this is the most challenging regime because the donor pool is systematically biased in ways that cannot be identified from observed data alone without extra assumptions or sensitivity parameters.}
\end{itemize}

\revtext{As the potential selection bias depends on the mechanism driving missing data, it is warranted for future work to conduct user studies to understand factors influencing voluntary data donation for fairness monitoring.}

\revtext{Several potential solutions could be developed for missing-robust fairness measurement:}

\begin{itemize}
    \item \revtext{\textbf{Reweighting and Post-stratification:} Donor records can be assigned weights so that their aggregate distribution matches trusted external or internal controls, such as known population marginals over observable features. This is especially useful under MCAR or MAR, and can be implemented in MPC through secure weighted sums and normalization without reconstructing individual sensitive attributes. Nonetheless, its accuracy highly depends on the quality and completeness of the external controls.}

    \item \revtext{\textbf{Imputation of Sensitive Attributes:} Missing sensitive attributes can be estimated using secure models trained on donor data. This gives more granular correction under MAR, but it must be used carefully because model parameters and highly predictive proxies may leak sensitive information if not tightly protected. For example, the model can be used only on-the-fly inside the protected computation to contribute to final group-level fairness aggregates, so that what is produced is the metric itself rather than a persistent individual-level imputation. Moreover, obtaining valid consent for data used in secure model training remains a practical challenge.}

    \item \revtext{\textbf{Expectation-based Estimation:} Instead of assigning each user to a single protected group, the protocol can use probabilistic group memberships such as $\eta_g(x)=\Pr(G=g\mid X=x)$ and compute fairness metrics in expectation. This is feasible under MAR when feature-conditioned priors are available, and it fits MPC well because the main computation reduces to secure multiply-and-sum operations. However, misspecified models may amplify bias, and similar to imputation, it carries a risk of sensitive information leakage through model inference if not properly protected.}

    \item \revtext{\textbf{Bounding and Sensitivity Analysis:} When MNAR is plausible and point identification is not credible, the protocol can report fairness intervals rather than a single estimate by considering worst-case or sensitivity-constrained assignments for non-donors. This is a conservative but robust option, making uncertainty explicit and avoiding overstating fairness conclusions when the missingness mechanism is fundamentally untestable. However, the resulting bounds may be too wide to support actionable decisions.}
\end{itemize}




\section{Industrial Dashboard and Lessons}
\label{appendixC}

\begin{figure}[t]
    \centering
\includegraphics[width=0.82\linewidth]{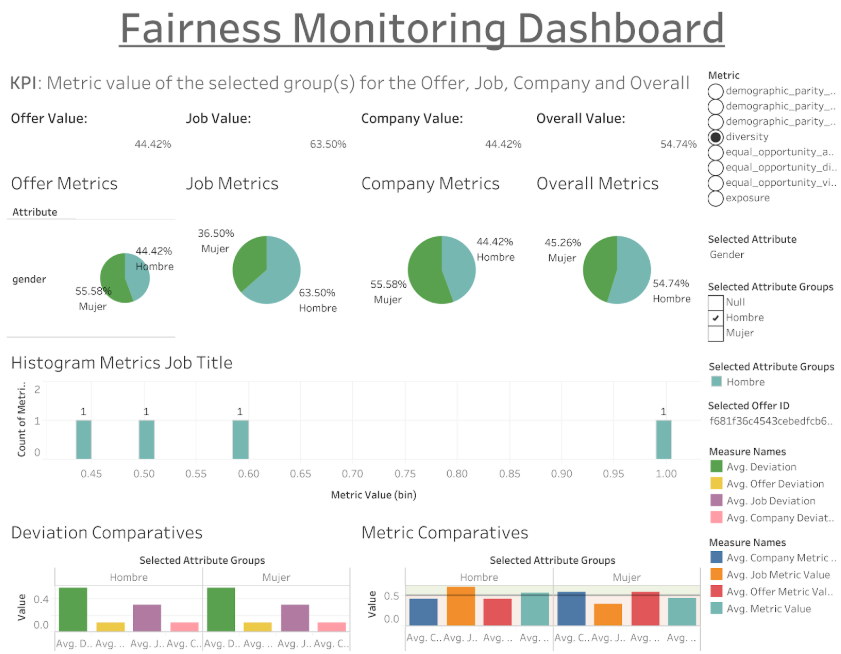}
    \caption{Industrial fairness monitoring dashboard implementing the proposed privacy-preserving protocol.}
    \label{fig:dashboard}
\end{figure}

Figure \ref{fig:dashboard} presents the industrial fairness monitoring dashboard implementing the proposed privacy-preserving protocol.
Here, we provide an illustrative use case to facilitate its interpretation. For instance, the dashboard may indicate that a given job offer has a female representation of $39.39\%$. While this value is close to the overall platform average of $43.30\%$, it is significantly higher than the $35.09\%$ average observed for comparable job title classifications. This contextual comparison enables recruiters and compliance stakeholders to assess how an individual offer performs relative to relevant benchmarks, rather than in isolation. Through the historical evolution view, recruiters can further examine whether female representation remains stable over time or improves following targeted interventions, such as distributing the offer to underrepresented audiences or revising job descriptions to be more inclusive. We provide an executive summary of organizational and technical lessons learned from the industry demonstrator, as shown in Table
\ref{tab:executive-lessons}.

\begin{table}[t]
\centering
\small
\begin{tabular}{p{0.38\linewidth} p{0.54\linewidth}}
\hline
\multicolumn{2}{l}{\textbf{1. Organizational and governance learnings}} \\
\hline
\textbf{AI governance and compliance foundation} &
The demonstrator translated abstract regulatory requirements into concrete monitoring practices, strengthening internal governance and readiness for AI Act compliance. \\

\textbf{Multidisciplinary approach to responsibility} &
Early and continuous involvement of technical, legal, product, and design roles proved essential to bridge gaps between research, regulation, and business practice. \\

\textbf{Data centrality and sensitive data dilemma} &
Fairness monitoring depends on high-quality protected attributes, yet legal and ethical constraints require pragmatic trade-offs to ensure deployability. \\

\textbf{Trustworthiness extends beyond algorithms} &
User trust is shaped not only by models and metrics but also by UI/UX, transparency mechanisms, and explicitly documented design choices. \\
\hline
\multicolumn{2}{l}{\textbf{2. Technical and implementation learnings}} \\
\hline
\textbf{Effort and development's cost} &
Most implementation effort was concentrated on data engineering, ETL pipelines, metric precomputation, and dashboard usability rather than metric formulation. \\

\textbf{Feasibility of discrimination monitoring dashboards} &
Privacy-preserving techniques such as secret sharing and secure multi-party computation were compatible with real production data and monitoring workflows. \\

\textbf{Complexity in interpretation and decision-making} &
The main challenge lay in enabling diverse stakeholders to correctly interpret metrics, particularly with multiple groups and intersectional analyses. \\
\hline
\end{tabular}
\caption{Executive summary of organizational and technical lessons learned from the industry demonstrator of an MPC-based fairness monitoring system.}
\label{tab:executive-lessons}
\end{table}

\end{document}